# Impact of dot size on dynamical characteristics of InAs/GaAs quantum dot lasers

Esfandiar Rajaei* and Mahdi Ahmadi Borji**

Department of Physics, The University of Guilan, Namjoo Street, Rasht, Iran

* Corresponding author, E-mail: Raf404@guilan.ac.ir, ** Email: Mehdi.p83@gmail.com

*Abstract:*

*The purpose of this research is to study laser dynamics of InAs/GaAs Quantum Dot Lasers (QDLs) by changing QD energy levels. To date, most of the investigations have focused on only one of these circumstances, and hardly the result of change in the energy levels can be seen in lasing. In this work, in the first step, energy levels of lens-shape QDs are investigated by the eight-band k.p method, their variation for different QD sizes are surveyed, and recombination energies of the discrete levels are determined. Then, by representing a three-level InAs/GaAs QD laser, dynamics of such a laser device is numerically studied by rate equations in which homogeneous and inhomogeneous broadenings are taken into account. The lasing process for both Ground State (GS) and Excited States (ES) was found to be much sensitive to the QD size. It was observed that in larger QDs, photon number and bandwidth of the small signal modulation decrease but turn-on delay, maximum output power, and threshold current of gain increase. It was also found that for a good modulation, smaller QDs, and form the point of view of high-power applications, larger QDs seem better.*

## I.    INTRODUCTION

Study of the structure of semiconductors enables to control their parameters such as energy gap, energy levels, band structures, etc; control of the basic factors of these structures results in high-performance devices. Semiconductor nanostructures include quantum dots (QDs), quantum wires (QWRs), and quantum wells (QWs) in which carriers are restricted in three, two, and one dimensions respectively. Quantum confined semiconductor nanostructures have been the focus of many researches due to their optical and electronic properties arising from quantum confinement of electrons and holes [1-4]. Progress in the fabrication of Quantum Dot (QD) lasers has recently attracted a huge attention to the application of quantum systems in optoelectronics [5-8]. In this article, we aim to study QD lasers which contain zero-dimensional semiconductors. These QD lasers are widely used in cable television signals, telephone and video communications, computer networks and interconnections, CD-ROM drivers, barcode scanning, laser printers, optical integrated circuits, telecommunications, signal processing, and a large number of medical and military applications.

Quantum dot laser (QDL) nano-devices due to the discrete density of states have a many good properties, namely, small threshold current, low temperature sensitivity, high optical gain and quantum efficiency, and high modulation bandwidth. Therefore, having a ubiquitous view of their energy states, carrier dynamics [9], and other physical features which affect the lasing process of a QD is instructive. Based on this fact, many research groups attempt to develop and optimize QDLs to fabricate optoelectronic devices with better performance [5, 10-13].

InAs/GaAs QDLs are mostly used in communication devices. Low threshold current density, high differential gain, and $1.3 \mu m$ lasing wavelengths have been observed in previous researches on GaAs based QD lasers. Gallium arsenide is a III-V direct band gap semiconductor which is crystallized in a zinc blend structure. It is usually used as substrate for the epitaxial growth of other III-V semiconductors such as InGaAs ternary alloy which itself has a direct band gap and is widely used in optoelectronics [14].

Self-assembled Stranski-Krastanov process is the common and efficient way of growing QDs. It can be performed by Metal Organic Chemical Vapor Deposition (MOCVD) or Molecular Beam Epitaxy (MBE) [15], in which dot density and size can be controlled. By changing the size and composition, QDs can be engineered for improved modal, differential, and total gain, modulation bandwidth, line-width enhancement factor, and for reduced threshold current. Thus, finding a way to enhance the efficiency of QDLs can be helpful. Among many materials, In(Ga)As/GaAs laser devices are focused by many scientists due to their interesting and applicable features [4, 10, 11, 16, 17].

Effect of change in the QD size can be so interesting but complicated, since many samples should be constructed and studied separately. Thus, very little experimental investigations are executed on this subject. So, numerical researches can be helpful in this situation. **Most of the investigations in this field have focused on only one of energy level change by size or lasing process, and hardly a paper can be seen studying the size variation effect in laser applications.** To date, many quantum solutions of Schrödinger equation are introduced, among which k.p approach with $8 \times 8$ matrix is a better approximation [18, 19].

The rest of this paper is organized as follows: section II is devoted to numerical calculation of QD energy levels and their behavior in different QD sizes by k.p model; in section III-A our three level model for laser dynamics is represented, and the results are discussed in III-B; finally, we make a conclusion in section IV.

## II. CONTROL OF ENERGY LEVELS

**Our model**

In growth of self-assembled InAs QDs on GaAs substrate, firstly a wetting layer including a few molecular layers is formed on the substrate [6]. After that, millions of quantum dots with random shape and sizes are formed on the wetting layer (WL) as a result of strain. The resulting system is usually covered by a GaAs cap-layer.

Many theoretical shapes can be approximated for QDs, namely, pyramidal, cubic, lens shape, cylindrical, etc. In this study we have considered lens-shape QDs. This assumption is a good approximation based on the former theoretical and experimental studies performed in references [6, 18, 20]. In this article, QDs are assumed to be far enough to disregard the quantum tunneling effects. Such calculations can be seen in projects previously done in [18, 19, 21]. The eight-band k.p approach was employed to solve the Schrödinger equation numerically in a self-consistent manner. We used the Dirichlet boundary condition for the electrostatic potential, and the strain effects were taken into account for a solution in real conditions. For a better performance of the simulation, a wide region is solved semi-classical but inside the dot and the WL that we aim to calculate their energy levels the solution is done with quantum mechanical approach [22].

Figure 1 illustrates the profile of a lens-shape InAs QD of diameter 20 nm, height of 7.5 nm, and a wetting layer of thickness 1.2 nm grown on the (001) plane of GaAs. In change of size, the height-to-diameter ratio remains fixed (i.e., $H = \frac{3D}{8}$ near the value ratio taken in the experimental work in [23]). The temperature is assumed to be $T = 300\ K$.

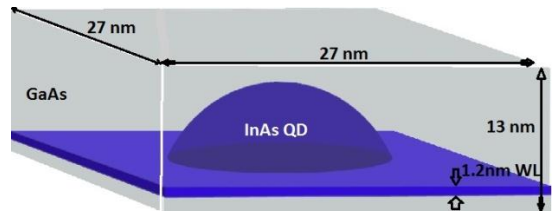

Fig. 1: Schematics of a lens-shape InAs QD with diameter 20 nm, height 7.5 nm and a WL of thickness 1.2 nm.

**Results for energy levels**

In Fig. 2 snapshots for the conduction and valence band-edges of InAs QD are shown in growth direction (i.e., z-axis) for three different sizes, beside which three lowest QD discrete energy states of electrons and holes are depicted.

As it is seen, by dimensional confinement of QDs, carrier energy bands are separated to atomic-like levels, whose difference increases for smaller QDs. For QDs of diameter less than 5 nm there found no electronic discrete level inside the QD. However, larger QDs appeared to have more separated electronic levels which have lower energies. In larger QDs, the discrete energy levels have become closer to each other, and they have been shifted to the bottom of the QD. On this figure, the Ground State (GS), and the Excited States ($ES_1$ and $ES_2$) as well as the relating stimulated recombination energies are schematically represented.

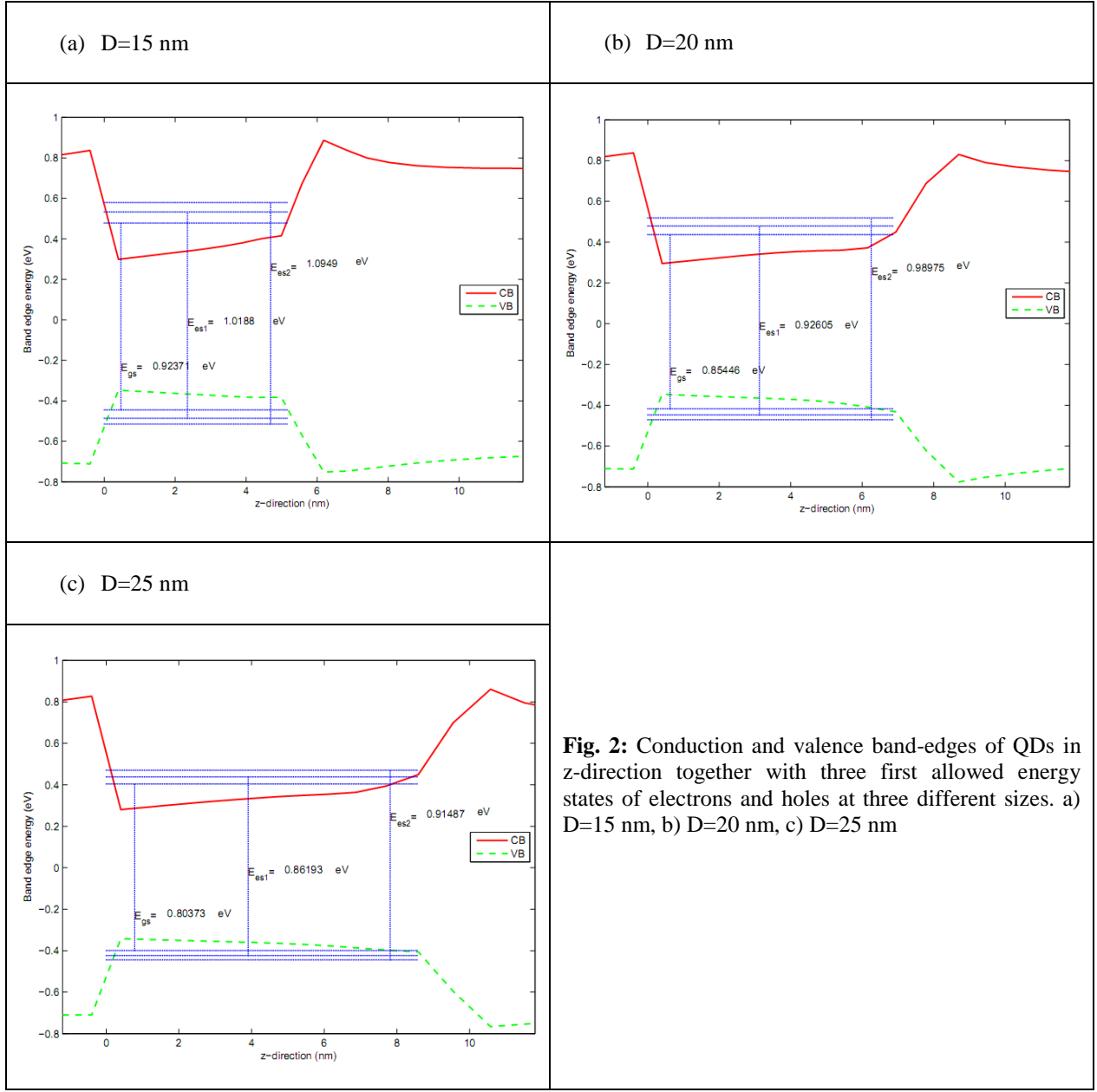

**Fig. 2:** Conduction and valence band-edges of QDs in z-direction together with three first allowed energy states of electrons and holes at three different sizes. a) D=15 nm, b) D=20 nm, c) D=25 nm

Energy gap in T=300K for bulk InAs is 0.36eV, and for GaAs is 1.43eV [24]. It has been shown that energy gap is sensitive to semiconductor size too. For instance, it is discussed in reference [20] that strain resulted from the 7% mismatch of lattice constants of InAs and its GaAs substrate is responsible for the change in the energy gap relative to its bulk sample. However, in our work, change of energy gap due to variation of QD size was negligible. It is due to the fact that energy gap can be very sensitive to the size only when the QDs are so small. It is proved in [25] that for QDs of diameter more than 6 nm there will be seen no remarkable dependence to size.

In Fig. 3 three lower energies of electrons and holes for various QD sizes as well as the energy of 2-molecular layer WL are depicted. As it is viewed, a larger QD gives rise to closer electron-hole (e-h) energy separation, which in turn leads to longer wavelength of photons from e-h recombination. Moreover, for this special WL thickness with substrate index (001), for diameters of less than 5 nm, there is no separate energy level inside the QD, and all the energy levels lay among the continuous GaAs energies; when D=5 nm, only GS level has come down into the QD, and for D=10 nm only GS and $ES_1$ levels appear in the QD. QDs of diameter more than 15nm include all GS, $ES_1$, and $ES_2$.

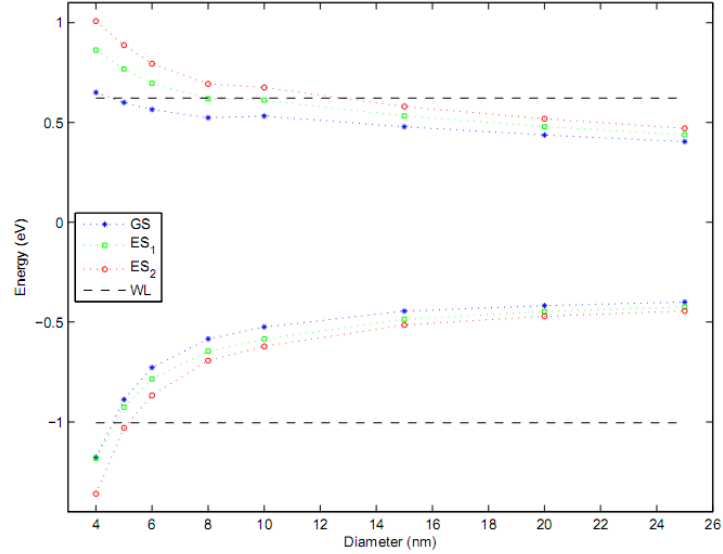

**Fig. 3:** GS, $ES_1$ and $ES_2$ energy levels for electrons and holes in different QD diameters. The fits are shown in dotted lines as a guide to the eye. Also, the WL energies are shown by dashed lines.

Also, change in e-h recombination energies, resulting from the first three eigenvalues of different QD sizes, are represented in Fig. 4. As it is seen, the recombination energies are reducing for greater QDs which lead to longer photon wavelengths in the laser. In addition, for larger QDs, the e-h recombination energy of GS and excited states are closer to each other and do not reduce with the same slope. These results show a good consonance with previously performed works in [26] on InAs/GaAs.

For a quantitative study of the behavior of energy states versus QD size, a cubic polynomial function was fitted to the data.

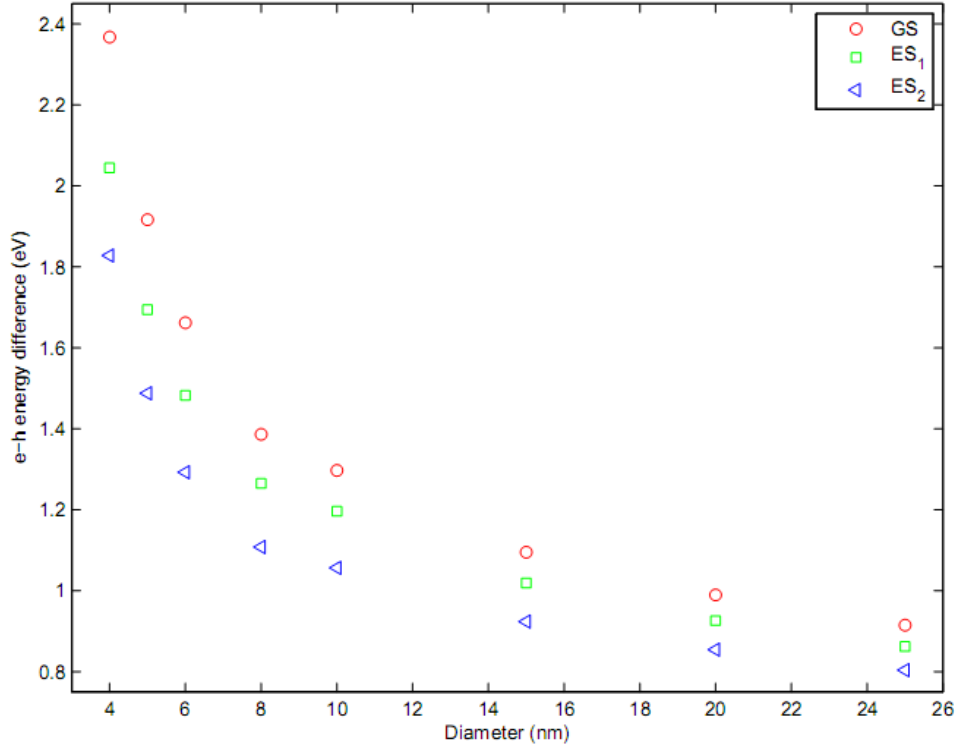

**Fig. 4:** Electron-hole recombination energy from GS, $ES_1$, and $ES_2$ at different QD sizes.

### III. LASER DYNAMICS

### III-A. Three level rate equation model

Fig. 5 illustrates the energy diagram of a three-level InAs/GaAs laser, including separate levels, namely, the GS, $ES_1$, and $ES_2$, and continuous level WL. At the beginning, a current is pumped into the WL, the major part of them are captured into $ES_2$ which can also relax into lower levels as well or even a few may return back to higher levels or directly recombine with holes and emit photons via stimulated emission. However, each of capture, escape, relaxation or recombination with holes, take some time which may lead to more frequent happening of a process relative to others. The required time for a process is dependent on the probability of occupation of origin and destination levels, requirements of Pauli Exclusion Principle, phonon bottleneck effect, etc. However, all these times are very short (~ps-ns) [27].

Taking into account all the transitions shown in the figure, and considering the homogeneous, and inhomogeneous broadening, and nonlinear gain the relating rate equations can be written as follows [ref]:

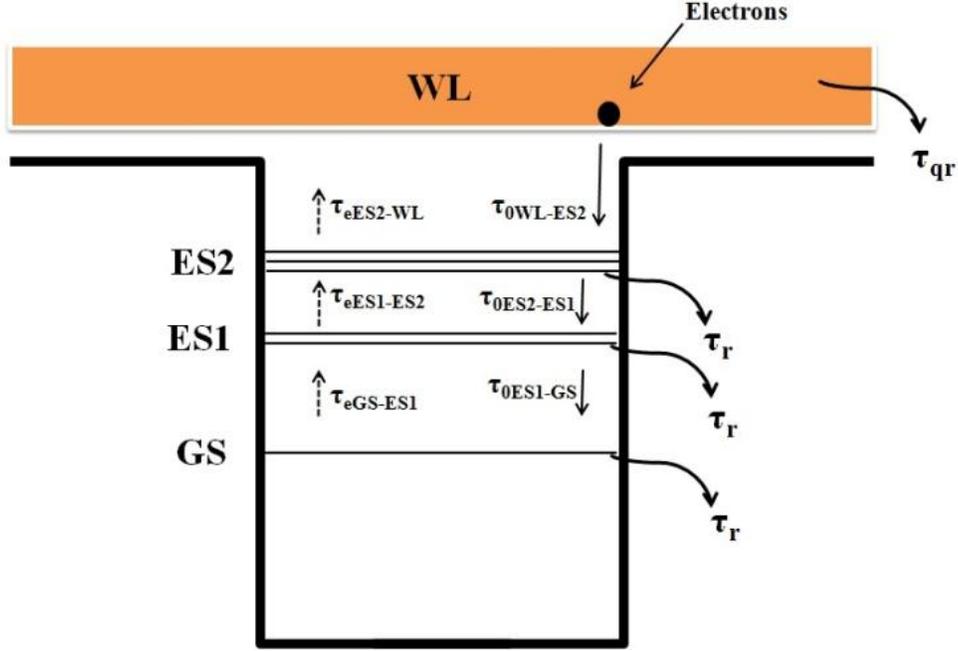

Fig. 5: Energy diagram of carrier transitions in three-level QD laser [ref].

$$\dot{N}_{WL} = \eta_i \frac{I}{e} - \frac{N_{WL}}{\tau_{qr}} + \frac{N_{ES2}}{\tau_{e_{ES2}}} - \frac{N_{WL}}{\tau_c}(1-f_{ES2}) \qquad (1)$$

$$\dot{N}_{ES2} = -\frac{N_{ES2}}{\tau_r} - \frac{N_{ES2}}{\tau_{e_{ES2}}} - \Gamma v_g K_{ES2}\frac{S_{ES2}(2f_{ES2}-1)}{1+\epsilon_{ES2}S_{ES2}} + \frac{N_{WL}}{\tau_c}(1-f_{ES2}) - \frac{N_{ES2}(1-f_{ES1})}{\tau_{0ES2ES1}} + \frac{N_{ES1}}{\tau_{e_{ES1}}}(1-f_{ES2}) \qquad (2)$$

$$\dot{N}_{ES1} = -\frac{N_{ES1}}{\tau_r} - \Gamma v_g K_{ES1}\frac{S_{ES1}(2f_{ES1}-1)}{1+\epsilon_{ES1}S_{ES1}} + \frac{N_{GS}}{\tau_{e_{GS}}}(1-f_{ES1}) - \frac{N_{ES1}(1-f_{GS})}{\tau_{0ES1GS}} + \frac{N_{ES2}(1-f_{ES1})}{\tau_{0ES2ES1}} - \frac{N_{ES1}(1-f_{ES2})}{\tau_{e_{ES1}}} \qquad (3)$$

$$\dot{N}_{GS} = -\frac{N_{GS}}{\tau_r} - \Gamma v_g K_{GS}\frac{S_{GS}(2f_{GS}-1)}{1+\epsilon_{GS}S_{GS}} - \frac{N_{GS}}{\tau_{e_{GS}}}(1-f_{ES1}) + \frac{N_{ES1}}{\tau_{0ES1GS}}(1-f_{GS}) \qquad (4)$$

$$\dot{S}_{ES2} = -\frac{S_{ES2}}{\tau_s} + \Gamma v_g K_{ES2}\frac{S_{ES2}(2f_{ES2}-1)}{1+\epsilon_{ES2}S_{ES2}} + \beta_{sp}\frac{N_{ES2}}{\tau_{sp}} \qquad (5)$$

$$\dot{S}_{ES1} = -\frac{S_{ES1}}{\tau_s} + \Gamma v_g K_{ES1}\frac{S_{ES1}(2f_{ES1}-1)}{1+\epsilon_{ES1}S_{ES1}} + \beta_{sp}\frac{N_{ES1}}{\tau_{sp}} \qquad (6)$$

$$\dot{S}_{GS} = -\frac{S_{GS}}{\tau_s} + \Gamma v_g K_{GS}\frac{S_{GS}(2f_{GS}-1)}{1+\epsilon_{GS}S_{GS}} + \beta_{sp}\frac{N_{GS}}{\tau_{sp}} \qquad (7)$$

where $\tau_s$ is the photon lifetime into the cavity, $\tau_{sp}$ is the spontaneous recombination time, $\tau_{0wl-ES_2}$ is the initial capture time to $ES_2$, and $\tau_{0ES_2-ES_1}$ and $\tau_{0ES_1-gs}$ are initial relaxation times respectively to $ES_2$ and GS. Also,

$$\tau_{ES1-gs} = \frac{\tau_{0ES_1-gs}}{1-f_{gs}}, \qquad \tau_{ES_2-ES_1} = \frac{\tau_{0ES_2-ES_1}}{1-f_{ES_1}} \qquad (8)$$

$\Gamma$ is the optical confinement factor, $\eta_i$ is the injection efficiency, $v_g$ is the group velocity into the cavity, and $\beta_{sp}$ is the spontaneous emission factor. $f_\alpha = N_\alpha/\mu_\alpha N_D$ is the occupation probability function in level $\alpha$ with degeneracy $\mu_\alpha$ and $N_D$ as the total number of QDs in the active region; $\epsilon_{m_\alpha}$ is defined as the gain compression factor of level $\alpha$

$$\epsilon_{m_\alpha} = \frac{e^2 p_{cv}^2 \tau_s}{4\hbar n_r^2 m_0^2 \epsilon_0 E_\alpha \Gamma_{hom}}, \tag{9}$$

with transition matrix

$$p_{cv}^2 = \frac{m_0^2 E_g (E_g + \delta)}{12 m_e \left(E_g + \frac{2\delta}{3}\right)}, \tag{10}$$

and gain factor $\epsilon_\alpha = \frac{\epsilon_{m_\alpha} \Gamma}{V_a}$ in which $\Gamma_{hom}$ is the homogeneous broadening factor. Also,

$$K_\alpha = \frac{2\pi e^2 \hbar \mu_\alpha \xi p_{cv}^2}{c n_r \epsilon_0 m_0^2 v_d \gamma_0 E_\alpha} \tag{11}$$

where $\gamma_0$ is the inhomogeneous broadening coefficient and $\xi = N_d V_d$ is the coverage of dots where $N_d$ is the dot density, and $V_D$ is the dot volume which is obtained as $V_D = 2\pi r^2 H/3$.

Here, $N_\alpha$ and $S_\alpha$ are respectively the carrier and photon numbers in energy level $\alpha$; we consider $E_g = 0.65 eV, \gamma_{hom} = 10 meV, \gamma_0 = 20 meV, \eta_i = 0.9$ and $\Gamma = 0.1$ [27]. For appropriate values of the variables introduced, we refer the reader to [27, 28]. These seven coupled differential equations have been solved simultaneously by the forth order Runge-Kutta method to achieve the lasing behavior in time.

**III-B. Results and discussions for the lasing process**

In this work, the dot density will be taken fixed. When the distance between QDs is variable due to QD size, but the dot density is fixed, it is expected to have the same volume densities of different-size-QDs in the fixed active region volume. The volume and surface for all QD sizes, were respectively $N_d = 1.055e23$ and $N_b = 1.37e15$, and the total number of QDs in the active region was $N_D = 8.23e7$.

Fig. 6(a) shows the GS photon number versus time when current I is 8 mA. As it is seen, initially, no photon is emitted but after an infinitesimal time which is named the turn-on delay, relaxation oscillations start which soon after that settle down to a stable radiation. The turn-on delay appears to be small for smaller QDs, but the difference is not more than a nanosecond. The delay is interpreted as the result of the time required for the injected carriers to be distributed in the active region, relax into the GS level, and be recombined with the holes. Therefore, some time is needed for the carriers to increase up to becoming sufficient for start of the simulated recombination [10]. The difference in turn-on delay for different sizes can be explained, by the fact that the energy difference between the barrier and the QD energy levels is less for smaller QDs. In addition, as it can be seen, the amplitude of relaxation oscillations decreases in larger QDs. Moreover, photon number in the steady state shows to be more for tiny QDs. This can be attributed to the threshold currents which are gain size dependent, as it is proved in our power-current curves.

Also, in Figs. 6(b, c) the photon number is plotted for lasing from $ES_1$ and $ES_2$. Energy of these levels is more than GS, and thus more current is needed for lasing from $ES_1$ and $ES_2$, since firstly lower levels use the current for lasing. By some trials the threshold current was found to be 1.4 A for $ES_1$ and 9.5 A for $ES_2$. As it is observed, the general behavior of turn-on delay and stable photon numbers is the same as for GS.

It is inferred from the figures that by size increase of QDs, the turn-on delay extends, and amplitude of relaxation oscillations and number of photons in the steady state decrease.

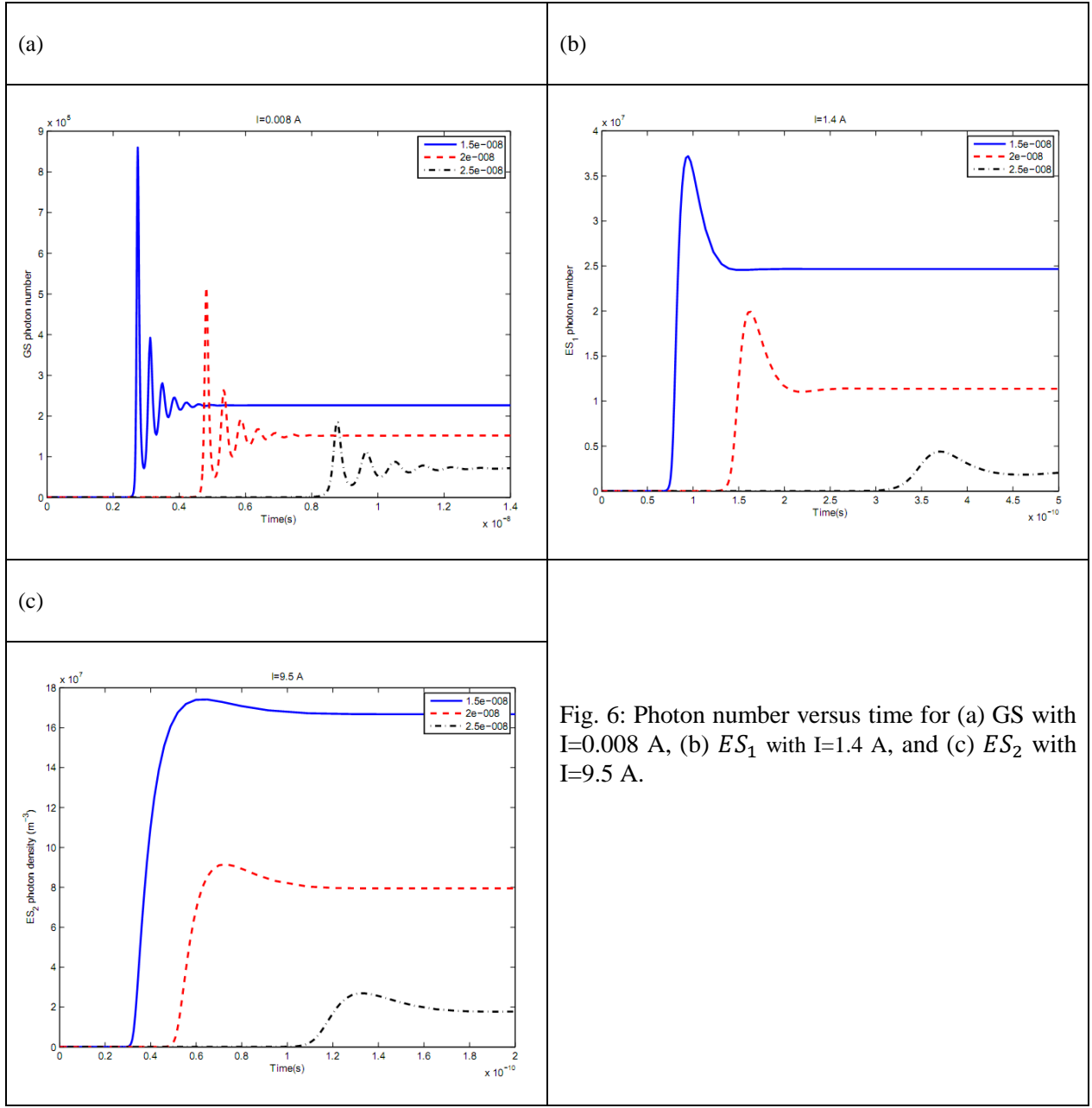

Fig. 6: Photon number versus time for (a) GS with I=0.008 A, (b) $ES_1$ with I=1.4 A, and (c) $ES_2$ with I=9.5 A.

The output power from level $\alpha$ is calculated as:

$$P_{out_\alpha} = \frac{cE_\alpha S_\alpha log\left(\frac{1}{R}\right)}{2n_r L} \tag{12}$$

in which $c$ is the light speed, R is reflectivity index, $n_r$ is the cavity refractive index, and $L$ is the cavity length. In Fig. 7 the output power is plotted versus current for GS, $ES_1$ and $ES_2$ for three different QD sizes. As it is seen, threshold currents of power are increased for larger QDs in all the levels. In addition, the power is more for smaller QDs, although the maximum power (which occurs in very high currents) is less for smaller QDs.

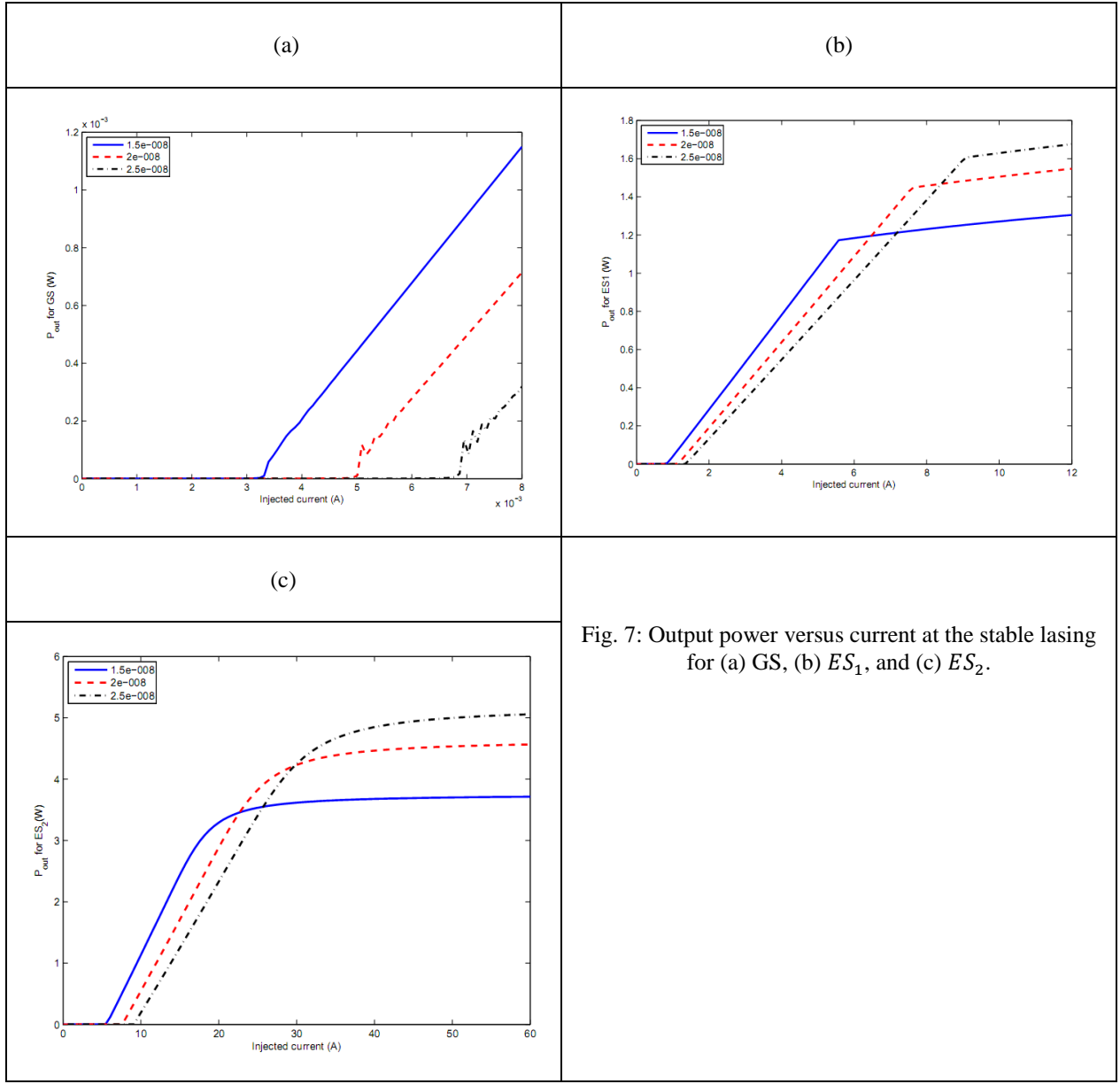

Fig. 7: Output power versus current at the stable lasing for (a) GS, (b) $ES_1$, and (c) $ES_2$.

The optical gain of level $\alpha$ is calculated by:

$$Gain_\alpha = \frac{\Gamma K_\alpha \left(\frac{2N_\alpha}{\mu_\alpha N_D} - 1\right)}{1+\epsilon_\alpha S_\alpha} \qquad (13)$$

In Fig. 8, laser gain is shown versus current for all the three energy states. As it is viewed, at weak currents, gain is negative for all levels. However, it increases to a positive value firstly for GS which grows to a saturation value at more currents. At higher amounts of current, $ES_1$ and $ES_2$ go to their positive maximum gain. Larger QD size, as it is observed, can enhance the threshold and saturation currents.

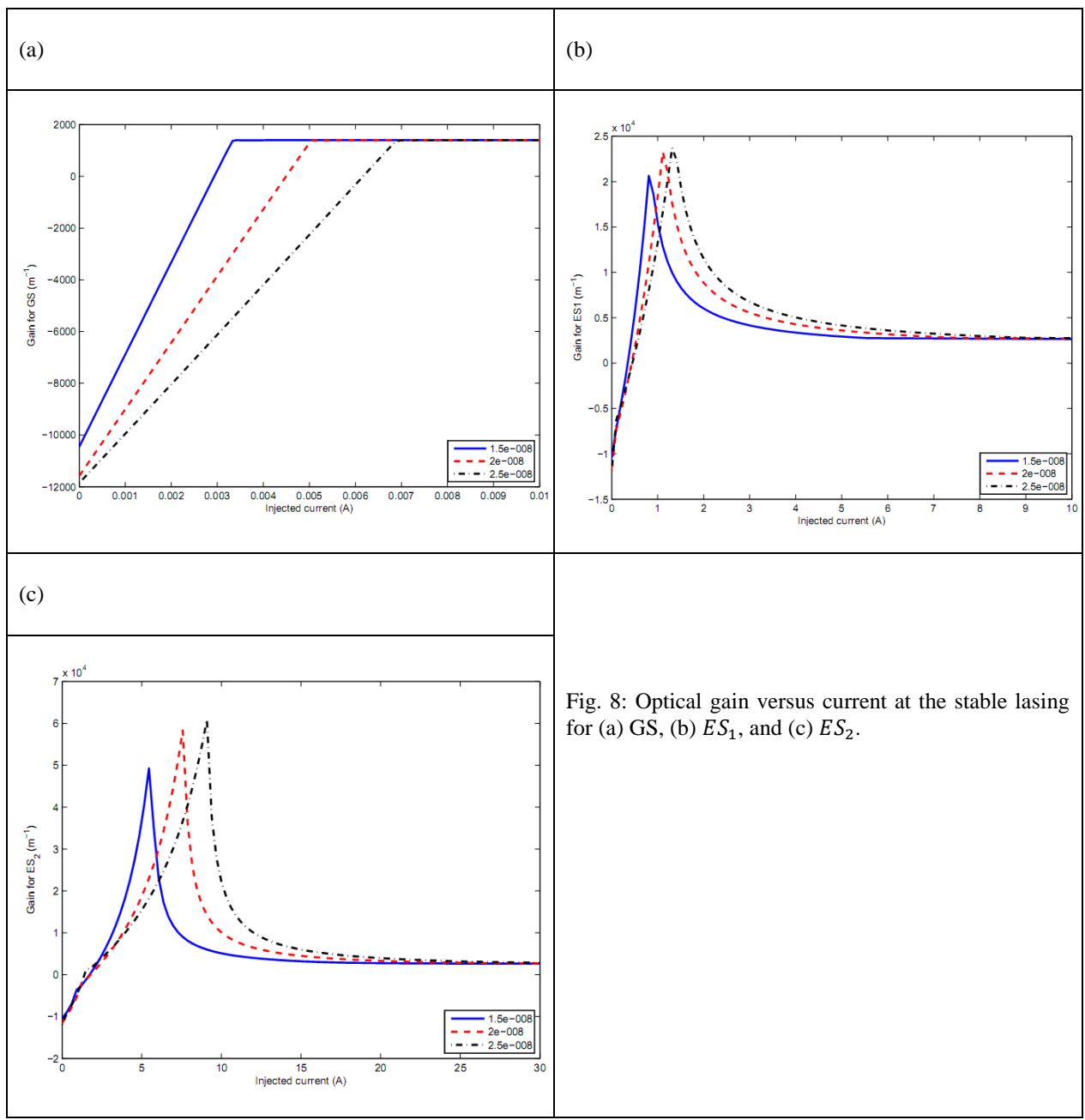

Fig. 8: Optical gain versus current at the stable lasing for (a) GS, (b) $ES_1$, and (c) $ES_2$.

In Fig. 9, also the modulation response [29] is shown in all levels for different QD sizes. It is viewed here that larger QDs lead to smaller modulation bandwidth.

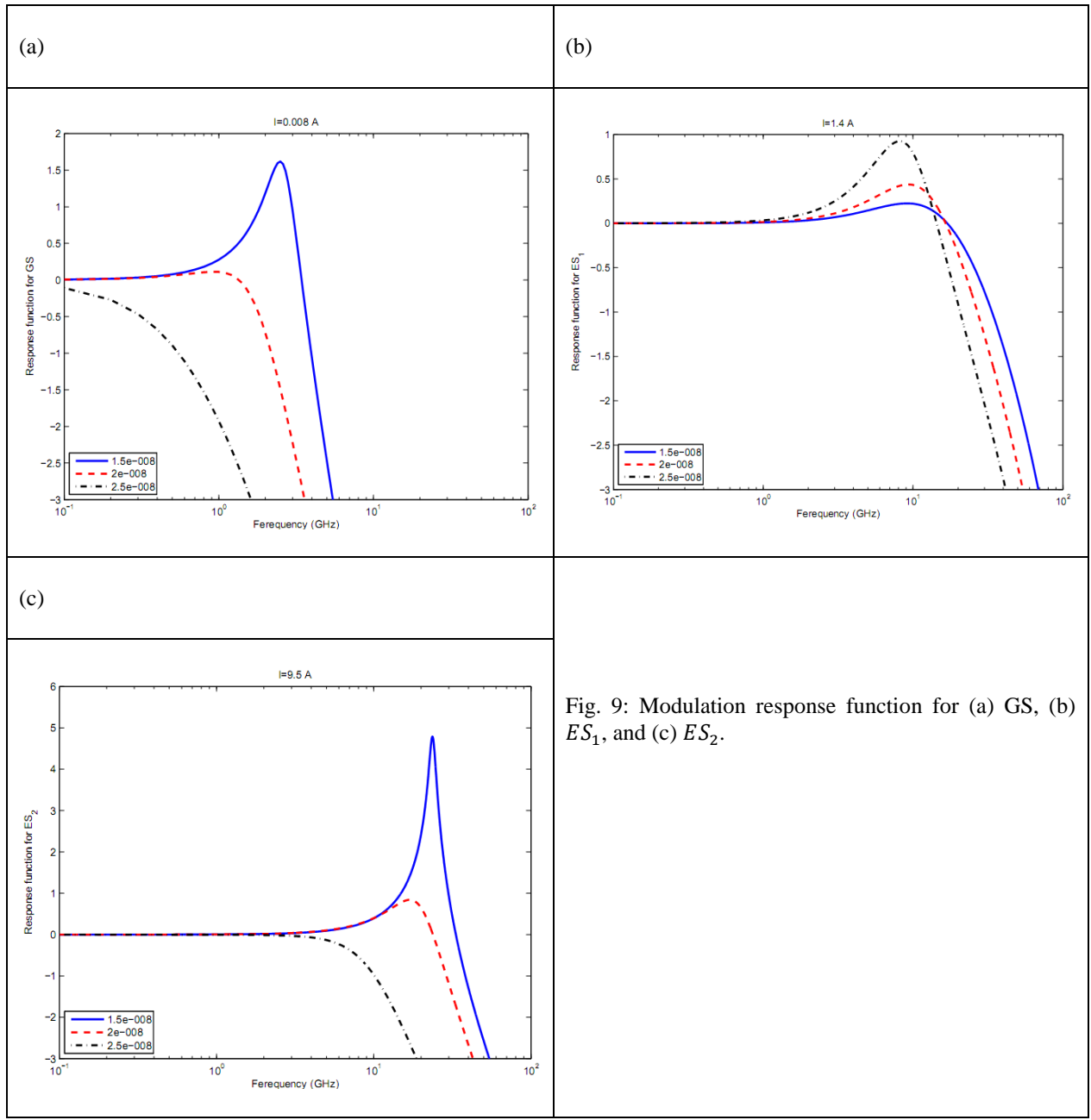

Fig. 9: Modulation response function for (a) GS, (b) $ES_1$, and (c) $ES_2$.

Fig 10 shows the behavior of output power in a wide range of values for inhomogeneous broadening and also at three different values of homogeneous broadening and QD sizes. From Fig. 10(a) it can be inferred that the trend is only descending for GS and no remarkable impact of QD size and homogeneous broadening can be viewed in the decrease of power in larger inhomogeneous broadening.

In Fig. 10(b&c) it is observed that in larger values of inhomogeneous broadening the power can be enhanced by increasing the homogenous broadening. That is the same for size increase as well.

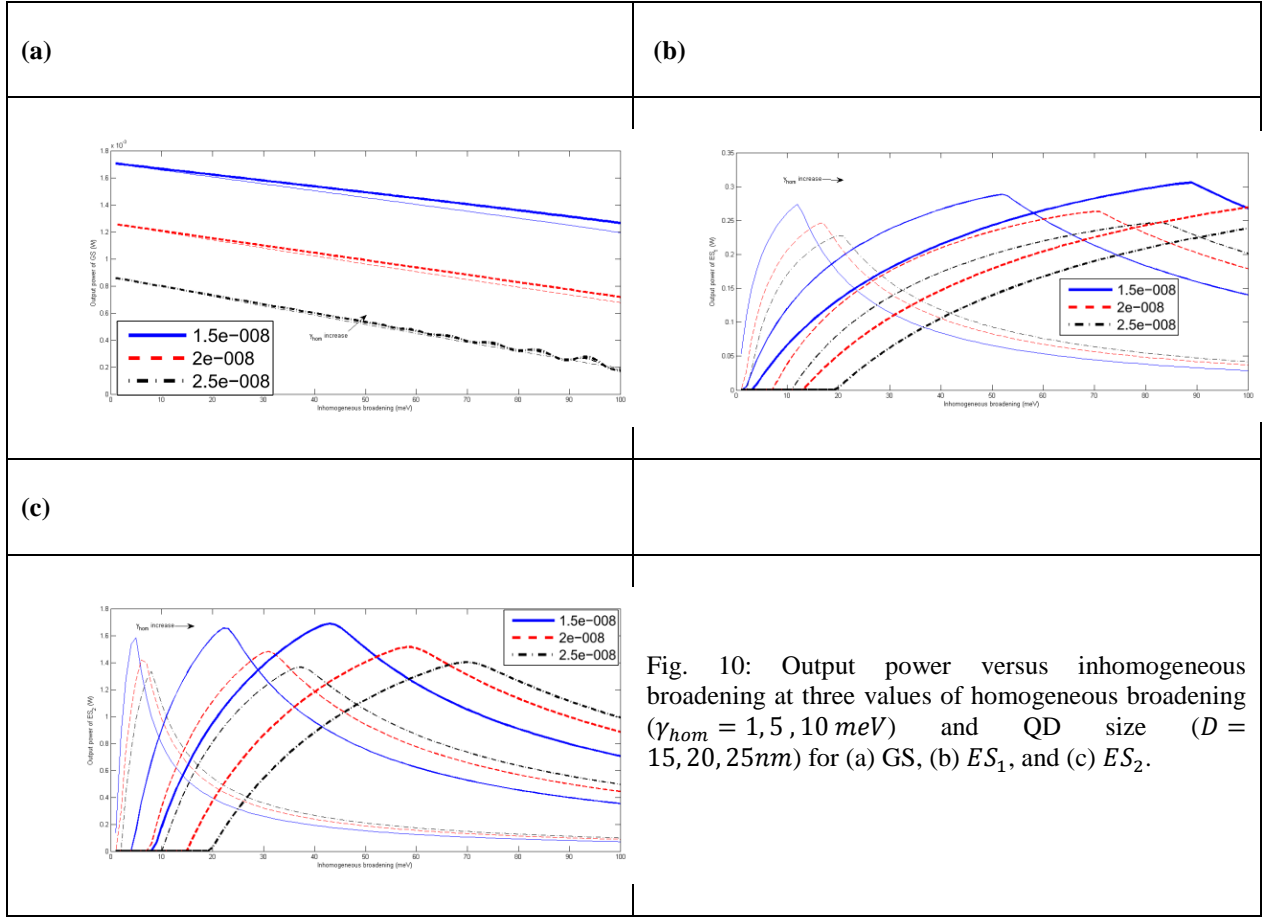

Fig. 10: Output power versus inhomogeneous broadening at three values of homogeneous broadening ($\gamma_{hom} = 1, 5, 10\ meV$) and QD size ($D = 15, 20, 25\ nm$) for (a) GS, (b) $ES_1$, and (c) $ES_2$.

## IV. CONCLUSION

We considered the band structure of lens-shape InAs QDs grown on GaAs substrate by k.p quantum solutions. Recombination energies of the discrete QD levels of different sizes were determined. The results showed a good consonance with Pryor et al results [21, 30]. After that, by a three-level laser dynamics and numerical solution of the rate equations, the laser properties were investigated. In larger QDs energy of the resulting photons decreased. Also, lasing outputs for all GS, $ES_1$, and $ES_2$ changed by QD size. It was found that in larger QD sizes, photon number and bandwidth of the small signal modulation decrease and turn-on delay, maximum output power, and threshold current of gain increase. In self-assembled QD lasers, in general, output power decreases by enhancement of inhomogeneous broadening, but as it was observed in our figures, for excited states, by increasing the QD sizes, larger output power can be achieved in higher values of inhomogeneous broadening. The small signal figures, proved that larger QDs result in the shorter modulation bandwidth and resonance frequency, and for all the levels, smaller QDs show a more bandwidth.

Therefore, from the point of view of modulation, smaller QDs, and from the point of view of high-power applications larger QDs seem better.

**Acknowledgement**

The authors give the sincere appreciation to Dr. S. Birner for providing the advanced 3D Nextnano++ simulation program [31] and his instructive guides. We would like to thank numerous colleagues, namely Prof. S. Farjami Shayesteh, Dr. S. Salari, K. Kayhani, and Y. Yekta Kia for sharing their points of view on the manuscript.

**References:**

1. Markéta ZÍKOVÁ, A.H., *Simulation of Quantum States in InAs/GaAs Quantum Dots.* NANOCON 2012. **23**(25): p. 10.
2. Ma, Y.J., et al., *Factors influencing epitaxial growth of three-dimensional Ge quantum dot crystals on pit-patterned Si substrate.* Nanotechnology, 2013. **24**(1): p. 015304.
3. DANESH KAFTROUDI, Z. and E. RAJAEI, *SIMULATION AND OPTIMIZATION OF OPTICAL PERFORMANCE OF INP-BASED LONGWAVELENGTH VERTICAL CAVITY SURFACE EMITTING LASER WITH SELECTIVELY TUNNEL JUNCTION APERTURE.* JOURNAL OF THEORETICAL AND APPLIED PHYSICS (IRANIAN PHYSICAL JOURNAL), 2010. **4**(2): p. 12-20.
4. Nedzinskas, R., et al., *Polarized photoreflectance and photoluminescence spectroscopy of InGaAs/GaAs quantum rods grown with As(2) and As(4) sources.* Nanoscale Research Letters, 2012. **7**(1): p. 609-609.
5. Bimberg, D., et al., *Quantum dot lasers: breakthrough in optoelectronics.* Thin Solid Films, 2000. **367**(1–2): p. 235-249.
6. Gioannini, M., *Analysis of the Optical Gain Characteristics of Semiconductor Quantum-Dash Materials Including the Band Structure Modifications Due to the Wetting Layer.* IEEE Journal of Quantum Electronics, 2006. **42**(3): p. 331-340.
7. KAFTROUDI, D., et al., *Thermal simulation of InP-based 1.3 μm vertical cavity surface emitting laser with AsSb-based DBRs*. Vol. 284. 2011, Amsterdam, PAYS-BAS: Elsevier. 11.
8. Asryan, L.V. and S. Luryi, *Tunneling-injection quantum-dot laser: ultrahigh temperature stability.* Quantum Electronics, IEEE Journal of, 2001. **37**(7): p. 905-910.
9. Horri, A., S.Z. Mirmoeini, and R. Faez, *Analysis of carrier dynamic effects in transistor lasers.* Optical Engineering, 2012. **51**(2): p. 024202-1-024202-6.
10. Yekta Kiya, Y., E. Rajaei, and A. Fali, *Study of response function of excited and ground state lasing in InGaAs/GaAs quantum dot laser.* J. Theor. Phys. , 2012. **1**: p. 246-256.

11. Azam Shafieenezhad, E.R., , Saeed Yazdani, *The Effect of Inhomogeneous Broadening on Characteristics of Three-State Lasing Ingaas/Gaas Quantum Dot Lasers.* International Journal of Scientific Engineering and Technology, 2014. **3**(3): p. 297- 301.
12. Saeed yazdani, E.R., , Azam Shafieenezhad, *Optimizing InAs/InP (113) B quantum dot lasers with considering mutual effects of coverage factor and cavity length on two-state lasing.* International Journal of Engineering Research, 2014. **3**(3): p. 172-176.
13. Asryan, L.V. *Dynamic characteristics of double tunneling-injection quantum dot lasers*. 2015.
14. Woolley, J.C., M.B. Thomas, and A.G. Thompson, *Optical energy gap variation in GaxIn1−x As alloys.* Canadian Journal of Physics, 1968. **46**(2): p. 157-159.
15. Gibson, R., et al., *Molecular beam epitaxy grown indium self-assembled plasmonic nanostructures.* Journal of Crystal Growth, 2015. **425**(0): p. 307-311.
16. Hazdra, P., et al., *Optical characterisation of MOVPE grown vertically correlated InAs/GaAs quantum dots.* Microelectronics Journal, 2008. **39**(8): p. 1070-1074.
17. Fali, A., E. Rajaei, and Z. Kaftroudi, *Effects of the carrier relaxation lifetime and inhomogeneous broadening on the modulation response of InGaAs/GaAs self-assembled quantum-dot lasers.* Journal of the Korean Physical Society, 2014. **64**(1): p. 16-22.
18. Zieliński, M., M. Korkusiński, and P. Hawrylak, *Atomistic tight-binding theory of multiexciton complexes in a self-assembled InAs quantum dot.* Physical Review B, 2010. **81**(8): p. 085301.
19. Korkusinski, M., M. Zielinski, and P. Hawrylak, *Multiexciton complexes in InAs self-assembled quantum dots.* Journal of Applied Physics, 2009. **105**(12): p. 122406.
20. Jiang, H. and J. Singh, *Conduction band spectra in self-assembled InAs/GaAs dots: A comparison of effective mass and an eight-band approach.* Applied Physics Letters, 1997. **71**(22): p. 3239-3241.
21. Pryor, C., *Eight-band calculations of strained InAs/GaAs quantum dots compared with one-, four-, and six-band approximations.* Physical Review B, 1998. **57**(12): p. 7190-7195.
22. Trellakis, A., et al., *The 3D nanometer device project nextnano: Concepts, methods, results.* Journal of Computational Electronics, 2006. **5**(4): p. 285-289.
23. Kamath, K., et al., *Small-signal modulation and differential gain of single-mode self-organized In0.4Ga0.6As/GaAs quantum dot lasers.* Applied Physics Letters, 1997. **70**(22): p. 2952-2953.
24. Bratkovski, A. and T.I. Kamins, *Nanowire-Based Light-Emitting Diodes and Light-Detection Devices With Nanocrystalline Outer Surface*. 2010, Google Patents.
25. Baskoutas, S. and A.F. Terzis, *Size-dependent band gap of colloidal quantum dots.* Journal of Applied Physics, 2006. **99**(1): p. 013708.
26. Jang, Y.D., et al., *Comparison of quantum nature in InAs/GaAs quantum dots.* Journal of the Korean Physical Society, 2003. **42**(Suppl): p. 111-113.
27. Lv, S.-f., et al., *Modeling and simulation of InAs/GaAs quantum dot lasers.* Optoelectronics Letters, 2011. **7**(2): p. 122-125.
28. Gioannini, M., *Ground-state power quenching in two-state lasing quantum dot lasers.* Journal of Applied Physics, 2012. **111**(4): p. 043108.
29. Horri, A. and R. Faez, *Small signal circuit modeling for semiconductor self-assembled quantum dot laser.* Optical Engineering, 2011. **50**(3): p. 034202-034202-5.
30. Pryor, C.E. and M.E. Pistol, *Band-edge diagrams for strained III\char21{}V semiconductor quantum wells, wires, and dots.* Physical Review B, 2005. **72**(20): p. 205311.
31. Birner, S., et al., *nextnano: General Purpose 3-D Simulations.* Electron Devices, IEEE Transactions on, 2007. **54**(9): p. 2137-2142.